\documentclass{JINST}
\let\ifpdf\relax

\title{Large scale Gd-beta-diketonate based organic liquid scintillator production for antineutrino detection}

\author{C.~Aberle, C.~Buck\thanks{Corresponding
author.}, B.~Gramlich, F.X.~Hartmann\footnote{current address: Prof. F.X.Hartmann, Hartmann Scientific, W.Lynn Shores Cir., City of Virgina Beach, Virgina, USA 23452-2609, email: hartmann.physics@cox.net.}, M.~Lindner, S.~Sch\"onert\footnote{now at Technische Universit\"at M\"unchen, D-80333 M\"unchen, Germany.}, U.~Schwan, S.~Wagner, H.~Watanabe\\
\llap{}Max-Planck-Institut f\"ur Kernphysik,\\
  Saupfercheckweg 1, 69117 Heidelberg, Germany\\

E-mail: \email{christian.buck@mpi-hd.mpg.de}}

\abstract{Over the course of several decades, organic liquid scintillators have formed the basis for successful neutrino detectors. Gadolinium-loaded liquid scintillators provide efficient background suppression for electron antineutrino detection at nuclear reactor plants. In the Double Chooz reactor antineutrino experiment, a newly developed beta-diketonate gadolinium-loaded scintillator is utilized for the first time. Its large scale production and characterization are described. A new, light yield matched metal-free companion scintillator is presented. Both organic liquids comprise the target and ``Gamma Catcher'' of the Double Chooz detectors.}

\keywords{Scintillators; Liquid detectors; Large detector systems for particle and astroparticle physics}

\begin{document}

\section{Introduction}

A critical issue for reactor antineutrino experiments is the need for optical and chemical stability of the gadolinium loaded liquid scintillator (Gd-LS), as learned from several past experiments~\cite{CHO, PV}. The Gd is used to detect neutrons following an antineutrino interaction with hydrogen nuclei. 

In the Double Chooz (DC) experiment \cite{DC} two new types of Gd-LS have been studied and further tested. These are \mbox{Gd-BDK\texttrademark} ~(for Gd beta-diketonate) and Gd-CBX\texttrademark ~(for Gd carboxylate)~\cite{TUM1, TUM2}. These two systems meet the basic requirements for the DC scintillator: chemical stability of the Gd molecules and the other LS components, compatibility with detector materials, transparency, intrinsic light yield, radiopurity, Gd solubility and the stability of these properties over several years of data taking. Optimization of the optical properties includes the need to tune the light yield while maintaining a constant density and matching the light emission to the spectral response of the photomultiplier tubes (PMTs).

The Gd-LS that is currently loaded in the DC Far Detector is based on beta-diketone chemistry, the first use in a large scale antineutrino detector. The chemistry of this new Gd-BDK scintillator \cite{FXHCB} is based in part on knowledge obtained from studies of the similar Indium (\mbox{In-BDK}) system \cite{InS} developed at MPIK, Heidelberg and used in the LENS (Low Energy Neutrino Spectroscopy) Prototype at Gran Sasso \cite{LENS}. The specific Gd organic scintillator candidate, that we initially considered, was based on the use of the simplest five carbon BDK anion -- ACAC (acetylacetonate). We found this BDK compound difficult to sublime and thus it lacked a potential productive route to achieving the level of optical and radiochemical purity needed in the DC experiment. Consequently, in the final version, the BDK was selected to be THD (2,2,6,6-tetramethyl-heptane-3,5-dionate) based on extensive research experience in producing solid, liquid and gaseous Gd-THD for testing in the \mbox{Ho-163} neutrino mass experiments \cite{Hol}. The more effective shielding of the metal ions by the ligands in the ``bulky'' THD version enhances its volatility and thermal stability since interaction between molecules and hydration effects are reduced compared to the ACAC complex~\cite{Eis}.  

The Gd-CBX scintillator is a pH-stabilized carboxylic acid scintillator system. It is based on the use of the trimethylhexanoic carboxylic acid, TMHA, and serves as a DC back-up. It was first created by us at MPIK, Heidelberg,~\cite{TUM2} with details presented in ref.~\cite{CBX2}. Studies of this system \cite{JPC} show that it does not have the molecular stability of the Gd-BDK system.

In the DC concept there are two detectors with identical design. One detector is positioned close (approx.~400~m) to the two nuclear reactor cores of Chooz. Its purpose is to measure with high precision the electron antineutrino rate coming from the reactor. A Far Detector at a distance of 1.05~km to the source is used to search for a disappearance signal of the electron antineutrino flux due to an oscillation effect. The Gd-LS for both detectors was prepared simultaneously and homogenized after production to assure the target liquids are identical at the near and far position. In each detector about 10.3~m$^3$ of Gd-LS are needed. Therefore we produced more than 20~m$^3$ of this liquid.

Reactor antineutrinos are detected via the captures on hydrogen nuclei in the scintillator molecules. In this reaction a positron and a neutron are created. The neutrons of the antineutrino reaction are mainly captured on the Gd. After neutron capture the excited Gd nucleus emits gammas with a total energy of about 8~MeV. The design of the DC Far Detector and the expected reaction rate in the target of about 70 antineutrinos per day determine the basic requirements on the liquids. The Gd-concentration in the target should be above 0.1~wt.\%. At such concentrations the fraction of neutron captures on Gd exceeds 80\% providing high detection efficiency. To keep the accidental background rate in DC below 1\% of the expected signal the allowed single rate in the neutrino energy window (0.7 - 12~MeV) is 10~Hz maximum~\cite{DC}. This constrains the allowed impurity levels in our scintillators to less than 10$^{-12}$~g/g for uranium and thorium and 10$^{-9}$~g/g for potassium. 

The design goal for the attenuation length of the target scintillator is above 5~m in the wavelength region from 430~nm to 500~nm. This is the region where the main scintillator emission occurs and also where the inner detector photomultiplier tubes are sensitive. Attenuation lengths significantly below 5~m would lead to inhomogeneities in the detector's light collection for the dimensional size of the detector. Here, the diameter of the cylindrical target vessel is 2.3~m and it has a height of about 2.6~m, thus absorption lengths greater than these dimensions are preferred. The optical properties have to be stable during the DC data taking period which is planned to be 5 years.        

The Gd-loaded DC target liquid (Target) is surrounded by an unloaded scintillator, the Gamma Catcher (GC). The purpose of the GC liquid in the DC experiment is to detect gamma radiation from the Gd deexcitation or positron annihilation which escape from the Target volume. The basic requirements for the GC liquid are similar to those for the Target. Some of the properties of the two scintillators have to be matched, e.g.~the densities have to agree within better than 1\% for mechanical reasons and the light yield for proper energy reconstruction of events. Other properties, such as the scintillation timing are tuned individually so as to allow for the localization of events by using pulse shape analysis. The GC liquids for the two detectors are mixed independently. The production of the GC for the DC Far Detector which was constructed and filled first is described. The total GC volume filled into the Far Detector is about 22~m$^3$.      

\section{Large scale scintillator production}

The scintillators for the inner DC detector were produced at MPIK Heidelberg using specially designed gas and liquid systems as well as dedicated tanks. A scintillator hall was specially designed for the production of the scintillators in DC but has general applicable features of interest to other experiments of this type. The hall consists of two parts. One half is a section containing a bay to handle three standard 20 foot isocontainers for shipment, temporary storage and processing of scintillator components. The second half is an area comprised of four tanks each holding half a loading of the DC inner detector volume. The hall is serviced by a high purity nitrogen system to provide for an inert atmosphere over the scintillator liquids as well as sparging and operating pressures for pumps and pneumatic valves. The systems in DC are all designed as modules. A scintillator fluid operating system handles the movement and purification of the liquids. A gas operating system handles the use of nitrogen gas. We use as well a mixing module and a weighing tank. The modules are transportable and used remotely at the DC filling site to accomplish transport and filling of the DC detector.

The chemical composition of the Target and GC liquids are summarized in table~1 where the CAS numbers refer to generally available materials. As indicated below, the materials we actually use were specially prepared by companies to achieve our radiopurity, optical purity and chemical stability requirements. The Gd-loading in the Target leads to a decrease in light yield compared to the metal free version~\cite{CA}. To match the light yield of the GC containing no Gd with the one of the Target the aromatic fraction in the liquid needs to be lowered. At the same time the density of the two liquids has to be kept the same to guarantee mechanical stability of the detector vessels. To meet all the requirements simultaneously the GC composition has to be significantly different from the Target and an additional solvent is needed. 

\sloppypar{The Gd compound of the scintillator was produced by Sensient Imaging Technologies, ChemiePark Bitterfeld, Wolfen; a company specializing in sublimation of volatile rare earth compounds in the plastic optics industry. Before full scale production the sublimation was first tested on smaller scale at our laboratories~\cite{FXHCB}. Few grams of the sample are introduced into the central region of a glass tube and heated to a controlled temperature of $150 - 200^\circ$C. A vacuum ($0 - 10$~mbar absolute pressure) is drawn at the exit of the tube while a trace flow of nitrogen or argon gas enters from the upstream side. The crystals of the purified Gd-complex condense at the unheated part downstream of the sublimation tube. After upscaling the sublimation at the company the purification rate could be increased to about 1~kg/day.}  

After production and sublimation of about 100~kg Gd(thd)$_3$, the powder (comprised of molecular crystals) was stored in the dark in an underground laboratory (shielding of about 15~mw.e.) and was contained in more than 100 dark brown glass bottles until the Target mixing started. All of the glass bottles were closed and sealed in airtight aluminized mylar foil bags under an argon atmosphere. The mixing steps described below were all done under an inert nitrogen atmosphere. All materials in contact with the liquids during Gd-LS production were either glass or fluorinated hydrocarbons: PVDF (polyvinylidene fluoride), PFA (perfluoroalkoxy fluoropolymer) or PTFE (polytetrafluoroethylene). A successful test run was made at the 200~liter scale to test the complete production chain before proceeding with the full scale scintillator production. 
    
\begin{table}[h]
\caption[Composition]{Composition of the Target and Gamma Catcher scintillators used in Double Chooz\label{DC-Scint}.}
\begin{center}
\begin{tabular}{lll}
Scintillator & Composition & CAS Number\\
\hline
Target & 80~\%$_\mathrm{vol}$ n-dodecane & 112-40-3\\
	& 20~\%$_\mathrm{vol}$ o-PXE (ortho-Phenylxylylethane) & 6196-95-8\\
	& 4.5~g/l~Gd-(thd)$_3$ (Gd(III)-tris-(2,2,6,6-tetramethyl-heptane- & 14768-15-1\\
        & 3,5-dionate)) & \\
	& 0.5~\%$_\mathrm{wt.}$ Oxolane (tetrahydrofuran, THF) & 1099-99-9\\
	& 7~g/l PPO (2,5-Diphenyloxazole) & 92-71-7\\
	& 20~mg/l~bis-MSB (4-bis-(2-Methylstyryl)benzene) & 13280-61-0\\ [5px]
Gamma Catcher & 66~\%$_\mathrm{vol}$ Mineral oil (Shell Ondina 909) & 8042-47-5\\
	& 30~\%$_\mathrm{vol}$ n-dodecane & 112-40-3\\
	& 4~\%$_\mathrm{vol}$ o-PXE (ortho-Phenylxylylethane) & 6196-95-8\\
	& 2~g/l  PPO (2,5-Diphenyloxazole) & 92-71-7\\
	& 20~mg/l~bis-MSB (4-bis-(2-Methylstyryl)benzene) & 13280-61-0\\
\hline
\end{tabular}
\end{center}
\end{table}

The first step in the Target production was to produce the Target concentrates. Oxolane (tetrahydrofuran, THF, Roth, $ > 99.9$\%, UV/IR-Grade) is added under nitrogen atmosphere to the Gd-$\beta$-diketonate compound (Gd(thd)$_3$) in the bottles at a ratio of 1:1 by weight. After a few days of stirring the Gd powder was liquefied completely by the oxolane. The liquid containing the dissolved Gd was then added to pure n-dodecane (Japan Energy) in a 600~liter PVDF mixing tank. This mixing tank is equipped with a nitrogen sparging ring, an additional mechanical stirring device, and sits on a balance so as to record the weights of each component. Typical Gd-concentrations in the resulting concentrates were 5 mg/g. 

In total there were 10 concentrates mixed during the large scale production with average weights of about 400~kg. Each concentrate was tested for transparency in 10 cm quartz cells using a UV/Vis spectrophotometer (Varian, Cary 400). The attenuation lengths at a wavelength of 430~nm in all concentrates (about a factor of four more concentrated than the Gd-concentration in the final mixture) were measured to be above 20~m, close to the limit of the instrument's sensitivity. 

In addition, each concentrate was checked for its water content, since the presence of water above the level of 50~ppm might reduce the scintillator stability. The water content was found to be below 35~ppm in all the samples. 

The concentrates, after testing, were added to a larger tank (10~m$^3$) that was situated on weighing sensors. The dimensions of this ``weighing tank'', made out of glass fiber reinforced plastic and an inner PVDF liner, were chosen so as to be able to hold the full Target volume for one detector. The concentrates already contain n-dodecane, so only the difference of the n-dodecane needed in the final Target loading and that already in the concentrates needs to be added as well to the weighing tank. 

In a next step, an aromatic solvent is added to the n-dodecane/Gd-BDK mixture in order to increase the light yield. For this purpose we use o-PXE, a molecule that fluoresces in the ultra-violet. Its density, near unity, is high enough above that of the n-dodecane that we can tune the density to the exact value needed throughout the detector. In addition to the o-PXE, two other fluors, PPO and bis-MSB, are added to shift the wavelengths of the o-PXE/n-C12 mixture to the region in wavelength matching the spectral response of the photomultiplier tubes. In this case, the PPO acts as a non-radiative transfer agent, transferring energy from the o-PXE molecules by Coulombic interactions to PPO, which itself emits photons. These photons are then wavelength shifted by the bis-MSB, whose concentration is held below that of the critical concentration.

The o-PXE isomer was specially produced by Dixie Chemicals, Houston, Texas, USA using the original catalytic procedure of Koch Industries. Close cooperation existed with the Dixie Chemcial Company to optimize the radiochemical and optical purity. Nonetheless, further optical purification was made. To improve the absorption length, a four-stage {\it Quadracolumn} purification technique was developed and used for the first time. The quadracolumn is based on the use of 3 differing pH activated column packings to remove acids, bases and neutralized particles; followed by a section to remove the column particles themselves. Thus, four layers of different absorber materials removed various optical impurities in the liquid. In total we purified 8 tons (for two detectors) of o-PXE in 42 purification runs. The attenuation length at 430 nm improved from 2.9~m after delivery to more than 8~m in the best purification runs. 

The PPO used in this experiment was specially prepared at Perkin-Elmer, Netherlands. Differing grades of the primary fluor from various companies were filtered (0.5 micron filter) and tested for transparency and radiopurity before the best PPO candidate was pursued. As a consequence, we arranged to produce the Perkin-Elmer PPO for the project (which henceforth was denoted as a new product, ``Neutrino Grade''). To achieve the purity goals of neutrino experiments an additional purification step was incorporated in the synthesis. After recrystallisation with an alcohol water mixture the company conducts an extra recrystallisation step for the Neutrino Grade PPO. Additional effective purification steps to remove potassium as water extraction of concentrated organic solutions or sublimation were not required. The secondary wavelength shifter, bis-MSB (Sigma-Aldrich) was used as is; it passed radio- and optical purity tests for our needs.

To complete the assembly of the liquids, several hundred liters of purified PXE were filled into the mixing tank. The PPO was then added at high concentrations up to 90~g/kg. Next, the bis-MSB was added to the PPO/o-PXE solution as a concentrate (200~mg/g). The subsequent fluor liquid, comprising a few hundred liters of the concentrated fluors (o-PXE/PPO/bis-MSB) is further tested and then moved into the weighing tank (already containing the Gd-solution, above) by passing through a 0.5 micron filter. Finally, pure o-PXE was added to the complete liquid mixture until the final design concentrations for all components were achieved. 

The sequence decscribed here, in which we first add the Gd-complex to n-dodecane, then prepare the concentrated fluors in PXE, mix, and add more PXE to achieve the final concentrations allows us to best be able to monitor the Gd concentration using UV/Vis methods as the UV absorption bands of the Gd-BDK are observable up until the addition of o-PXE and fluors. 

The preparation of the 23~m$^3$ GC liquid was directly done in a 20~ft stainless steel transport tank in the MPIK scintillator hall. The two main components in this scintillator mixture are a mineral oil, Ondina 909 from Shell, and n-dodecane, as used in the Target. Of all the mineral oils tested, Ondina 909 showed best performance in terms of transparency due to its low aromaticity. As for the Target, the PPO for the GC was first dissolved in pure o-PXE at a high concentration and then added to the transport tank after testing and filtering. The bis-MSB was added directly as a powder to the transport tank. The nitrogen system in our scintillator hall was used to nitrogen flush the tanks and keep them on a nitrogen blanket. The pressure reducer of the low pressure nitrogen system (LPN) is used to keep the pressure in the tanks at a level of 30~mbar above atmospheric pressure during operations. At an overpressure of more than 50~mbar a relief valve at the end of the LPN line starts to blow-off nitrogen gas. In this way the pressure is kept in a defined range during the production of the liquids. 

The  liquids were transported from MPIK to the EdF nuclear reactor site in Chooz, France, for transfer into the temporary storage and filling system of the DC Far Detector. The GC was transported in the isocontainer in which it was prepared. The Target liquid is transferred using a specially made 6 ton PVDF transport tank (using two shipments). The Target liquid for a Near Detector in Chooz was transferred into two 5~ton PVDF temporary storage tanks for future use (pending completion of the DC Near Detector).   

\section{Scintillator properties}

\subsection{General properties}

The densities of the DC liquids all have to agree to within at least 1\%. Otherwise there is a mechanical risk of fracturing the cylindrical detector vessels. This is one of the main reasons why we have chosen multiple component liquids in the four volumes of the DC detector, as this choice allows for the tuning of the density as well as the light yield (seen below). 

The DC liquids all have identical densities of $0.804\pm 0.001$~kg/l as measured at 15$^\circ$C. The densities and kinematic viscosities serve as important input parameters into the design of the liquid systems. The kinematic viscosity of the Target was measured to be 2.32~mm$^2$/s at 21$^\circ$C. Since mineral oil is more viscous than n-dodecane, a higher value of 3.7~mm$^2$/s at a similar temperature was found for the GC. 

During all stages of the production, the water content of the Target was determined by coulometric Karl-Fischer Titration (Aqua 40.00, AnalytikJenaAG). The Target is a hygroscopic liquid and thus much work was made to limit the exposure of the scintillator to water. The presence of water in Gd scintillators can lead to decomposition of the Gd complexes by hydrolysis, this leads to a deterioration of the optical properties and ultimate ``collapse'' due to precipitation. A final Target sample was taken in the Far Detector neutrino laboratory before filling. The water concentration was measured to be $36\pm1$~ppm within our specification of less than 50~ppm. The GC liquid is less hygroscopic than the Target and its stability is not as sensitive to water. 

\subsection{Optical properties}

\begin{figure}[tb]
\begin{center}
	\includegraphics[scale=0.50]{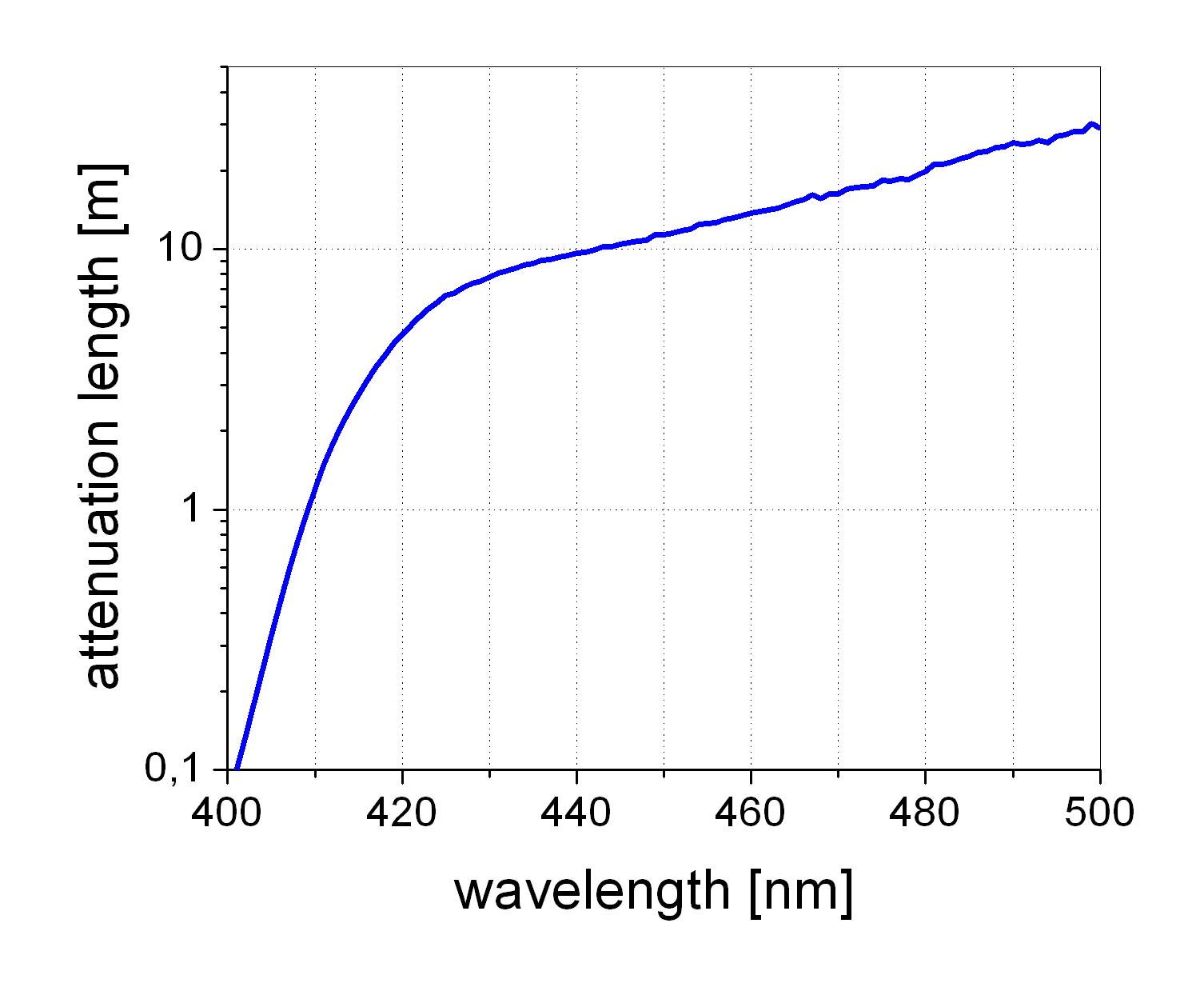}
\end{center}
\caption[]{Target attenuation length as measured before the Double Chooz Far Detector filling. 
\label{Fig1}}
\end{figure} 

In figure \ref{Fig1}, the attenuation length of the Target is shown in our wavelength region of interest. A sample taken out of the weighing tank at the Far Detector site in the neutrino laboratory at Chooz, before the final filtering stage (0.5~$\mu$m pore size) and detector filling, was determined in a 10~cm cell using a UV/Vis spectrophotometer. As reference a blank measurement of the empty cell is used to zero the baseline~\cite{JL}. Corrections for reflection losses at the cell windows are applied due to different refractive indices in air and the liquids. The correction includes the wavelength dependence of the refractive index in the quartz material of the cell. The attenuation length in the Target was determined to be $7.8\pm0.5$~m at 430~nm which is well above our design goal of 5~m. Within the errors of the measurements this number is consistent with what is expected based on the calculated value using the molar extinction coefficients of the summed individual components. These absorbance measurements were made for each component. In the spectrophotometer, absorbed light is not distinguishable from scattered light. In the larger DC-like detector any scattered light can still be detected, barring absorption, by the photomultiplier tubes.

\begin{figure}[tb]
\begin{center}
	\includegraphics[scale=0.50]{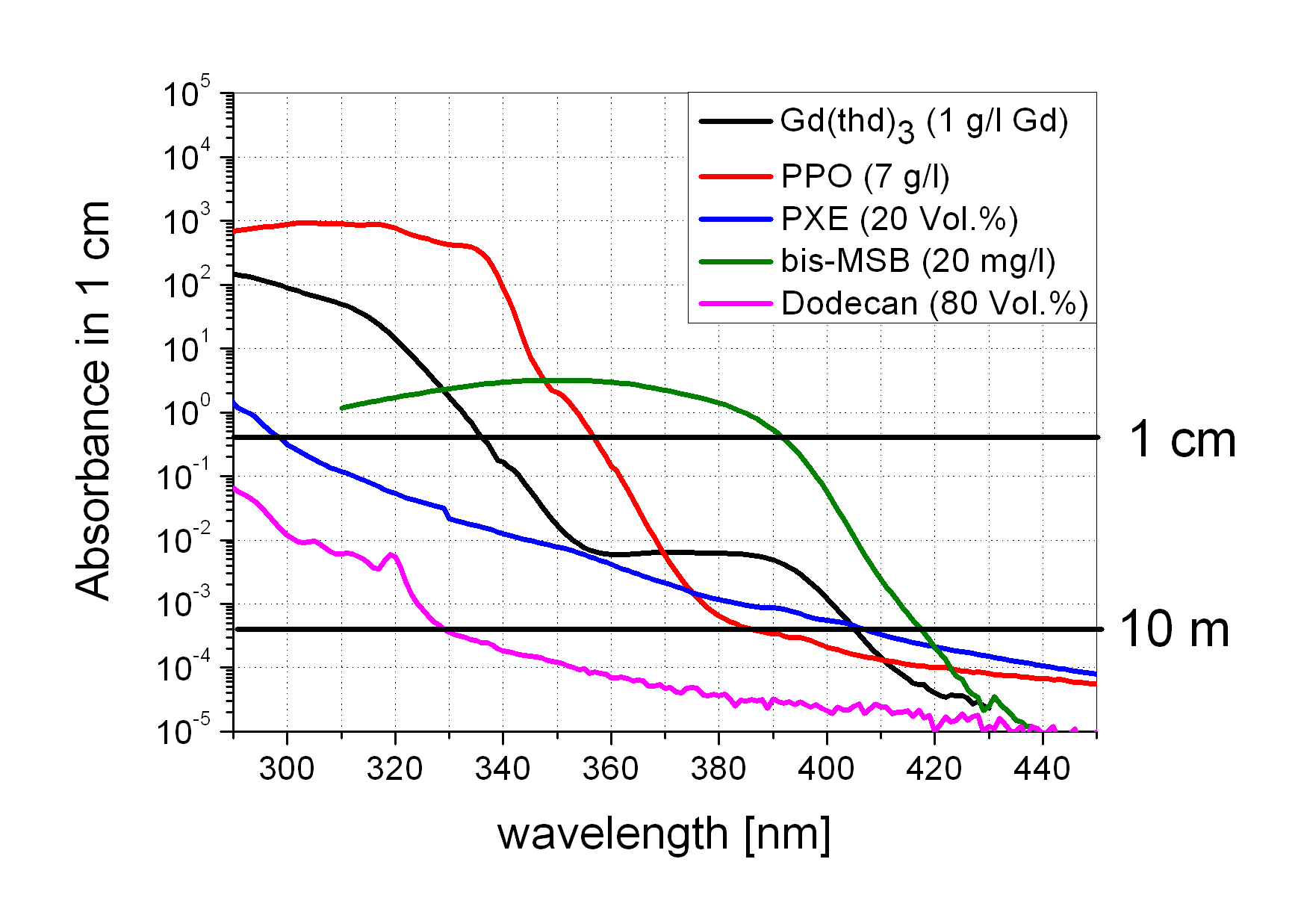}
\end{center}
\caption[]{Contribution of the Target components to the absorbance of the full mixture. The curves are on the left side of the plot from top to bottom for PPO, Gd, PXE, Dodecan (at 290 nm) and bis-MSB (starting at 310 nm).
\label{Fig2}}
\end{figure}

The primary scintillation light, after non-radiative energy transfer and wavelength shifting by the fluors, is produced at wavelengths between 350~nm and 500~nm. The absorbance of the Target scintillator is dominated at longer wavelengths than $425$~nm by the absorption of the \mbox{o-PXE} and PPO. In this region the main part of the absorbance is most likely due to impurities in these chemicals which are assumed in the worst case scenario to not re-emit photons. Figure~2 shows the contribution of each Target component to the absorbance. As seen in this figure, the bis-MSB absorption dominates at wavelengths smaller than 420~nm. At these wavelengths the primary scintillation light is significantly re-emitted by the bis-MSB, as the bis-MSB has a quantum yield of 0.94~\cite{Ber}. Since the photons are re-emitted at longer wavelengths, the emission spectrum becomes shifted on the distance that the scintillation light traveled through the liquid. The greater the distance, the more the spectrum is shifted to longer wavelengths due to the bis-MSB.   

\begin{figure}[tb]
\begin{center}
	\includegraphics[scale=0.50]{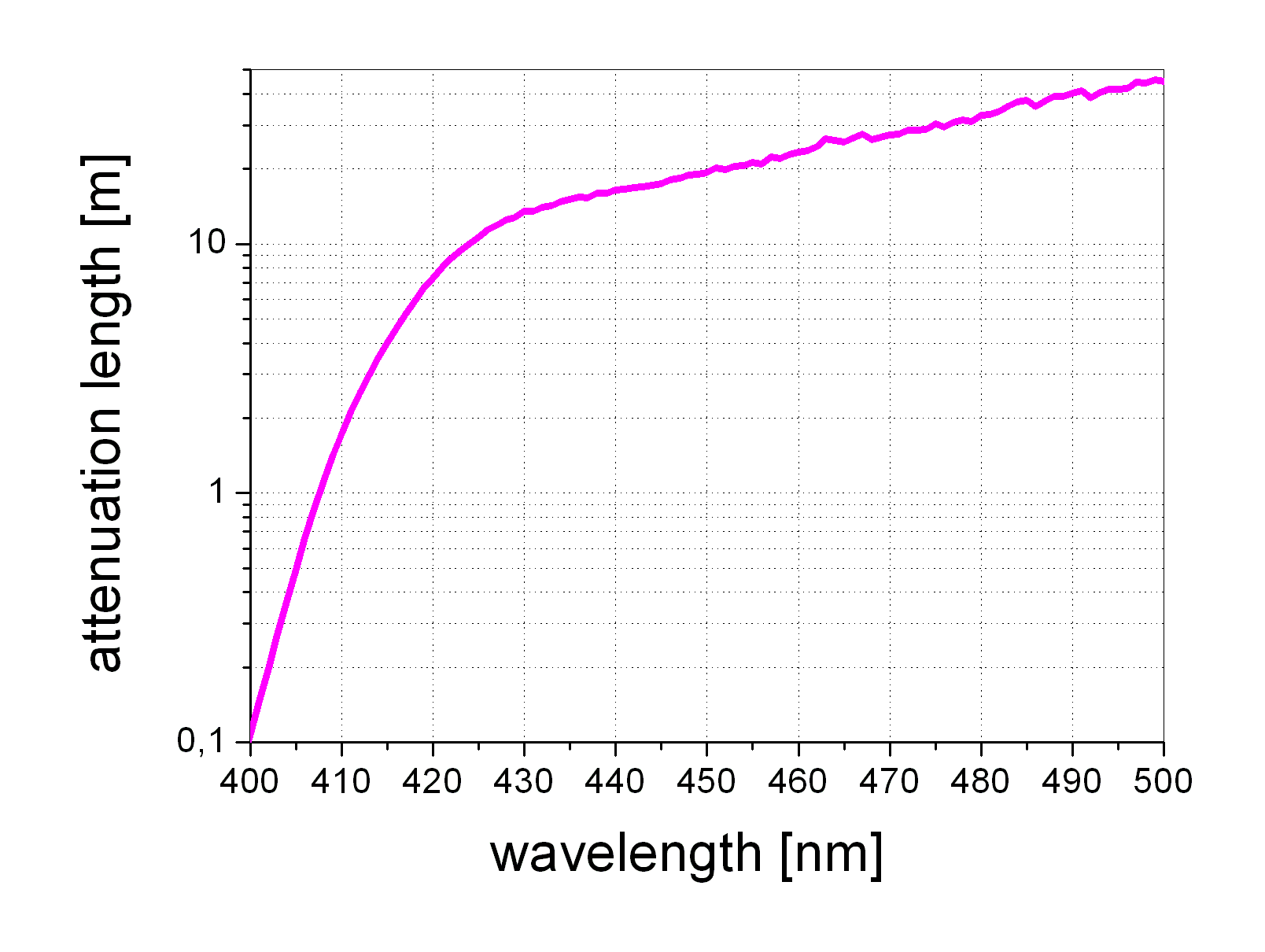}
\end{center}
\caption[]{Gamma Catcher attenuation length as measured before the Double Chooz Far Detector filling.  
\label{Fig3}}
\end{figure}

Figure 3 shows the attenuation length measured for the DC Gamma Catcher scintillator. The sample was taken after the liquid transport to Chooz. The attenuation length is above 10~m in the most relevant wavelength region. This liquid is more transparent than the Target since it contains much less PPO and o-PXE which are the limiting components as regards the absorbance in the optical region.  

The Target light yield was measured with an experimental set-up that compares the Compton backscattering peak (using a $^{137}$Cs gamma source) of the sample to those of several standards~\cite{CA}. Compared to an o-PXE scintillator using similar fluor amounts and containing no Gd nor n-dodecane the Target light yield was measured to be $57.6\pm1.0$\%. Both the Gd compound as well as the dilution due to n-dodecane serve to reduce the light yield of the scintillator. The n-dodecane is needed in the scintillator to ensure material compatibility with the acrylic Target vessel, as without it, the o-PXE would dissolve into it and soften it. 

The GC light yield was matched to that of the Target. In order to do this, the light yield in the GC was reduced by lowering the o-PXE and PPO concentrations (thus compensating for the light quenching caused by the Gd-molecules in the Target). Model predictions based on energy transfer rates were used for this tuning of luminescent properties \cite{CA}. For geometrical reasons the light collection efficiency in the GC volume is slightly above that for the Target volume. Therefore we adjusted the GC light yield to 97\% of the Target value to optimize the homogeneity of the overall light collection in the detector. Light yield quenching parameters for low energy electrons \cite{SW} and alpha particles \cite{CAD} were determined.

The PPO concentration in the GC was optimized for timing properties to allow separation of Target and GC events by pulse shape analysis. The scintillation time is slower at low PPO concentration. In dedicated laboratory measurements the scintillation photon emission time profiles were determined for the Target and for several GC candidates meeting the light yield matching requirements~\cite{CAD}. A start-stop technique with two PMTs in a common setup was used to get the time profiles of the samples. In one PMT the excitation time is determined whereas the second PMT measures single photons of the same scintillation event at low probability. The time differences of the two PMT signals are distributed as the probability density function for the scintillation emission time. To enhance the difference of the timing behavior in the two DC detector regions a GC candidate with substantially lower PPO concentration compared to the Target was chosen for the DC experiment.

\begin{figure}[tb]
\begin{center}
	\includegraphics{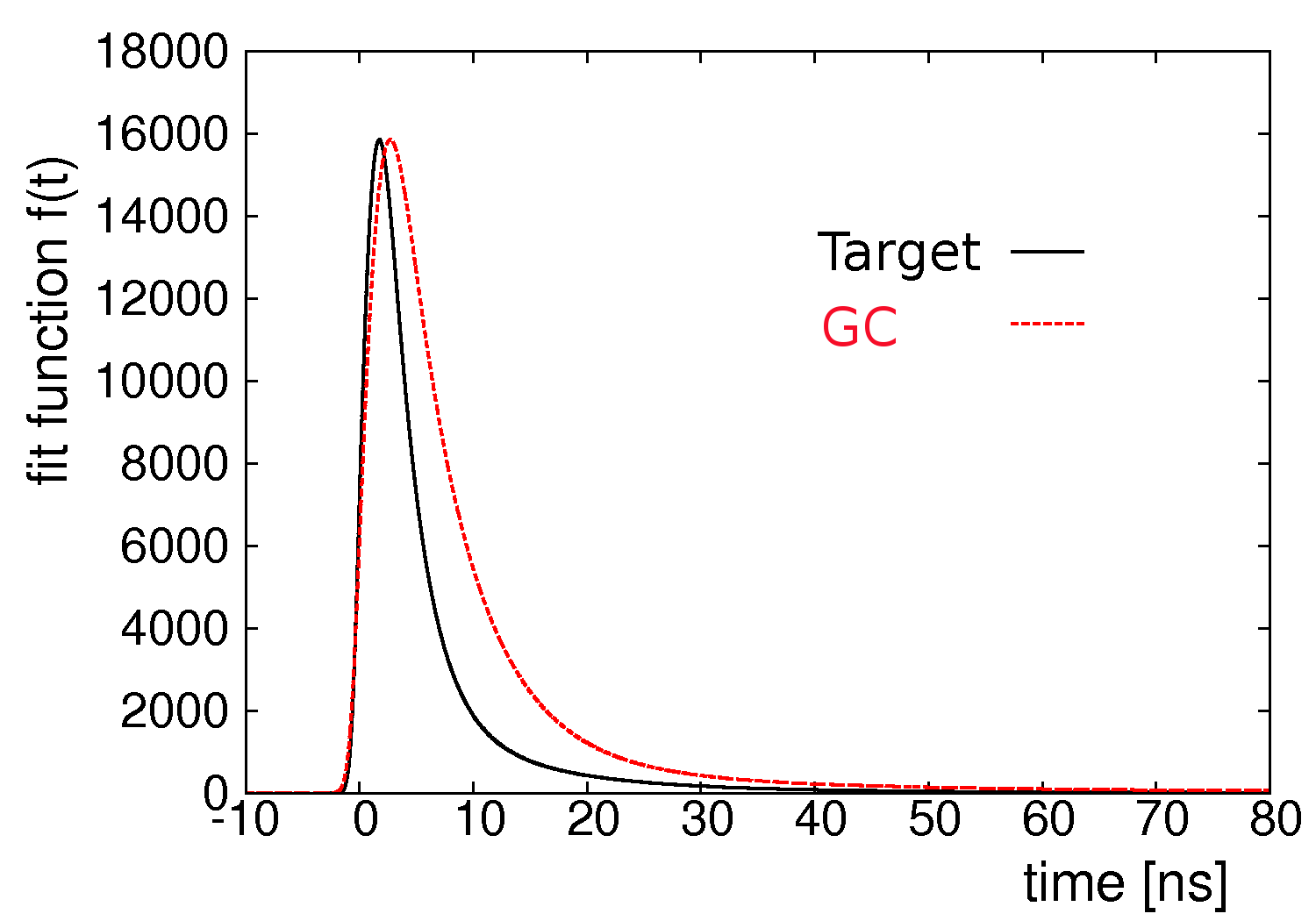}
\end{center}
\caption[]{Scintillaton photon emission time distribution for beta excitation in Target and Gamma Catcher.
\label{Fig4}}
\end{figure}

In Figure \ref{Fig4} the fits to the measured scintillaton photon emission times are shown for Target and GC. The timing behavior can be mathematically described by the sum of three exponentials with three timing constants $\tau_i$ and three weighting factors $q_i$. Due to the low PPO concentration the measured time constants in the GC ($q_1 = 0.81$, $\tau_1= 5.4$~ns, $q_2 = 0.12$, $\tau_2= 17$~ns, $q_3 = 0.07$, $\tau_3= 57$~ns) are higher than the ones of the Target ($q_1 = 0.71$, $\tau_1= 2.6$~ns, $q_2 = 0.25$, $\tau_2= 9.7$~ns, $q_3 = 0.04$, $\tau_3= 38$~ns). 

Monte Carlo simulations have shown that this discrepancy can be used to discriminate between energy depositions in the Target or GC volume of the DC detector. This could be useful to improve the knowledge on a certain class of events in the DC detector where the neutrino interacts in the GC and the created neutron is captured on Gd inside the Target after travelling through the acrylic vessel~\cite{CLD}. The uncertainty of the simulation for these so-called ``Spill In'' events is rather high. If these events could be determined with higher precison it improves the sensitvity of the DC experiment during its first phase with one detector taking data.

\subsection{Radiopurity}
Careful material selections were made, and dedicated purification steps were carried out, so that we could achieve the low-level radiopurity requirements for the scintillator chemical components. The main source of radioimpurities in the scintillator mixture is expected to be the powders (fluors and Gd-complex) as these are not purified by distillation, such as for the organic liquids, and their basic chemistry applicable to their production is conducive to picking up contaminating ions such as K$^+$. In addition, the organic solvents that were chosen were already tested for radiopurity in previous low energy neutrino experiments. The n-dodecane used in DC is from the same plant as the one used in the KamLAND experiment. They demonstrated the radiopurity of this liquid in terms of U, Th and K~\cite{KAM}. The o-PXE was investigated in the Borexino counting test facility (CTF) and the level of radioimpurities found \cite{PXE} was well below the DC specifications. 

In order to achieve the required radiopurity of the components, a major effort was made to eliminate the impurities in the Gd powder. Metal-$\beta$-diketone complexes are known for their stability and high vapor pressure. These features allow one to purify the 100~kg of Gd(thd)$_3$ by sublimation, thus removing significant amounts of the impurities since they are left behind as non-volatile inorganic species or themselves sublime under different conditions of temperature and pressure. The purification of the Gd-complex was done in close collaboration with Sensient Imaging Technologies. Whereas the starting material before sublimation contained potassium up to more than 100 ppm, it could be reduced in the sublimation process to less than 1~ppm. In total we measured 14.23~kg of sublimed Gd(thd)$_3$ divided in 5 samples and counted using low-background gamma-ray spectroscopy. For this purpose we used our GeMPI detector~\cite{Heu}, which is the world's most sensitive Ge-detector located at the Laboratori Nazionale del Gran Sasso (LNGS), Assergi (AQ), Italy. The only isotope which was above the detection limit for all the samples was $^{40}$K. The average activity was 13.5 mBq/kg corresponding to 0.43~ppm of natural potassium. This translates to an activity of the Target in one DC detector of about 0.6~Bq without any energy cut. This number has to be compared with an expected trigger rate due to total internal and external radioactivity of 5 - 10 Hz above threshold. The results of the Ge-spectroscopy are listed in table~\ref{radioGd}. Evidence for the presence of $^{152}$Eu was found in part of the samples. This isotope can be produced by cosmic-ray activation. From the measured activity no notable contribution to the background of DC is expected.      

\begin{table}[h]
\caption[radioGd]{Radiopurity results for the Gd(thd)$_3$ powder as measured by gamma-spectroscopy and their contribution to the Gd-scintillator. For potassium, the weighted average of all samples is given.\label{radioGd}}
\begin{center}
\begin{tabular}{lcc}
Isotope & concentration in Gd-complex & concentration in Target\\ \hline
$^{nat}$K & 0.43~ppm & 2.4~ppb \\
 $^{235}$U & $< 1$~ppb & $ < 6$~ppt\\ 
$^{238}$U ($^{226}$Ra) & $<0.05$~ppb & $< 0.3$~ppt\\
$^{228}$Th & $< 0.2$~ppb &  $< 1.1$~ppt \\ \hline \\
\end{tabular}
\end{center}
\end{table}

The PPO samples that we received from companies contained significant amounts of potassium. The potassium arises due to the presence of potassium containing compounds used in the synthesis of the PPO and is thus expected.  Perkin-Elmer (Netherlands) developed a new synthesis procedure that improved the removal of potassium and also took care to use procedures that would lower the other common radioisotopes mentioned above. The final PPO that was used in the scintillator production had potassium concentrations of 0.25 -- 0.5~ppm (7 - 14~mBq/kg) as summarized in table~\ref{radioPPO}. The measured value varied in the given range possibly due to differences in the production batches, different systematics of the analysis methods or inhomogeneities of the distribution of impurities in the powder. 

Three methods were used to detect potassium in this primary fluor powder: atomic absorption spectroscopy (in the MPIK chemistry laboratory), gamma-spectroscopy (with our GeMPI detector at LNGS, Italy) and neutron activation analysis (measured at TU M\"unchen, Germany). The total amount of PPO in one inner DC detector is about 120~kg. Therefore the expected activity in a DC detector due to the PPO contribution would be $0.8 - 1.7$~Bq before any energy cut, and thereby acceptable within limits. 

The bis-MSB concentration in the scintillators is rather low and therefore the constraints on its radiopurity are not as critical as for the other components. Nevertheless, it was tested using Ge-spectroscopy and found to be within specifications as well. 

\begin{table}[h]
\caption[radioPPO]{Radiopurity results for PPO (Perkin-Elmer, Neutrino Grade) measured by NAA, AAS and \mbox{$\gamma$-spectroscopy}. For the NAA (Neutron Activation Analysis) and the AAS (Atomic Absorption Spectroscopy) the average value of three measurements is reported. The mass of the sample counted using gamma-spectroscopy in the MPIK GeMPI detector (located in the underground laboratory at Gran Sasso, Italy) was 1300~g. \label{radioPPO}}
\begin{center}
\begin{tabular}{lcc}
Isotope & method & concentration in PPO \\ \hline
$^{nat}$K & NAA & $0.35\pm0.16$~ppm \\
$^{nat}$K & AAS & $0.25\pm0.04$~ppm \\
$^{nat}$K & GeMPI & $0.5\pm0.1$~ppm \\
$^{235}$U & GeMPI & $< 0.72$~ppb\\ 
$^{238}$U ($^{226}$Ra) & GeMPI & $< 0.05$~ppb\\
$^{228}$Th & GeMPI & $< 0.26$~ppb \\ \hline \\
\end{tabular}
\end{center}
\end{table}

Since potassium was expected to be the main radioimpurity in the liquid scintillators, we tested the final mixtures for this element using AAS. Only upper limits on the potassium concentrations could be obtained. The expected concentration of natural potassium in the Target and the GC, separately, due to the contributions of PPO and Gd(thd)$_3$ was $4 - 7$~ppb and $0.6 - 1.4$~ppb respectively. In the AAS measurement an upper limit of $^{nat}$K$<2$~ppb was obtained for both liquids. We note that the upper limit for the concentration of the final Target mixture is below the expected contributions from the powders. This observation can be explained assuming that impurities in the powders which contain the main part of the potassium do not fully dissolve in the scintillator solvent and were filtered out during scintillator production. 

\subsection{Number of hydrogen nuclei}

For a precision measurement on the reactor antineutrino flux as planned in DC an accurate knowledge on the number of Target $^1$H nuclei (called ``proton number'') is mandatory. In the period when both DC detectors are operational this systematic error cancels since the Target liquids are chemically identical. However, DC started data taking with a Far Detector only. The proton number in the Target is calculated using the weight measurements of the neat, chemically pure components during scintillator mixing. Table~\ref{weights} shows the measured weights of the Target components for the two production runs NU-01 and NU-02. The weight fraction of hydrogen atoms in the scintillator molecules is calculated to be 13.60$\%$ in each of the two cases corresponding to $8.12\cdot10^{25}$ hydrogen atoms per kg of the Target liquid. 

\begin{table}[h]
\caption[weights]{Weights and purity of the Target components used in the production runs NU-01 and NU-02. Numbers in parentheses are the result of tests as reported by the manufacturer using gas chromatography. The numbers for oxolane and bis-MSB are specifications given by the manufacturer. \label{weights}}
\begin{center}
\begin{tabular}{lcccc}
Component & purity [\%] & wt.~NU-01 [kg] & wt.~NU-02 [kg] & H [wt.\%] \\ \hline
n-dodecane & [99.1~\% (99.5~\% n-alkane)] & 6481 & 6482 & 15.39\\
o-PXE & [99.2~\% (99.6~\% PXE)] & 2130 & 2131 & 8.63\\
Oxolane & $ > 99.9$~\%   & 49 & 49 & 11.2\\
Gd(thd)$_3$ & sublimed & 48.6 & 48.6 & 8.5\\
PPO & neutrino grade & 75.5 & 75 & 5.0\\
bis-MSB &$ > 99$~\% & 0.22 & 0.22 & 7.1\\
\hline \\
\end{tabular}
\end{center}
\end{table}

The error on the proton density in the liquid is dominated by the precision of the knowledge of the concentrations of the dominant components, n-dodecane and o-PXE. The errors on the weights given in table~\ref{weights} for these two liquids are estimated to be 10~kg. 

The n-dodecane has a purity of 99.1\%. From the test report provided by Japan Energy, it is known that there are 0.2\% n-tridecane (C-13) and 0.2\% n-undecane (C-11) molecular species in the liquid. These additional n-alkanes (n-paraffins) have a negligible effect on the proton density. Thus 0.5\%, more or less, of unidentified components remain. These correspond to about 30~kg in one Target. 

For the case of the aromatic liquids in the scintillator mixture, it is composed mainly of PXE isomers at the level of 99.6\% PXE of which the main part is ortho-PXE. The remainder (about 10~kg per Target batch) of the aromatics is comprised of mainly styrene dimers (about 0.3\%).     

To determine an upper bound on the hydrogen fraction we assume there is 10~kg more of \mbox{n-dodecane} (high H concentration) and 10~kg less PXE (low H concentration) in each Target than measured. In addition, the assumption is made that all THF (low H concentration), which itself has a rather high vapor pressure, was lost by evaporation. In such a conservative scenario, the hydrogen fraction rises by about 0.15\% (relative) to 13.62\% of the total mass. This increase could be slightly higher in the case that the impurities of the solvent have a higher hydrogen content than n-dodecane. Such an effect, though, would be expected to be much smaller than the increase calculated above. 

For the lower bound on the hydrogen fraction we raise the amount of PXE by 10~kg and reduce the n-dodecane content by the same number. Furthermore we assume that the impurities in the chemicals have on average only half of the hydrogen content of the pure component. This is a rather conservative approach since most of the impurities are similar hydrocarbons themselves; in which case, the ratio of hydrogen to carbon is essentially unchanged. The total amount of water in the scintillator is measured to be less than 0.5~kg and therefore has no significant effect. With the above assumptions we get a hydrogen ratio of 13.56~wt.\%. The relative error on the hydrogen ratio in the Target scintillator is thus estimated to be relative +0.15\% / -0.3\%. Collecting terms, the amount of hydrogen by weight in the Target scintillator is thus (13.60$^{+0.02}_{-0.04}$\%) $^1$H by weight.

For the GC scintillator, it was not useful to calculate the proton number based on the weights of the ingredients since the hydrogen fraction in the mineral oil is not known to a good enough precision for our purposes here. Therefore, the hydrogen fraction in the GC liquid was measured using CHN analyses (Carbon, Hydrogen, Nitrogen) performed at BASF company, Ludwigshafen. The hydrogen content was found to be $14.6\pm0.2$~wt.$\%$. The improved $^1$H weight uncertainty of the Target over that of the GC shows the better performance of using actual total weights over that of CHN chemical analyses of small samples. This is achieved by using Target components that are pure, ``neat'' liquids, i.e.~of well-defined molecular weight and correspondingly avoiding the use of mixtures, such as most mineral oils. 

\begin{table}
\caption[Properties]{Main properties of the Target and GC scintillators used in Double Chooz. As a standard for the light yield measurement we used the liquid scintillator BC-505: Bicron, St. Gobain Crystals (80\% light yield anthracene). \label{Sumprop}}
\begin{center}
\begin{tabular}{lcc}
 & Target & Gamma Catcher\\ \hline
Density [kg/l] at 15$^\circ$ & $0.8035\pm0.0010$  &  $0.8041\pm0.0010$ \\ 
kinematic viscosity [mm$^2$/s] at 21$^\circ$ & $2.32\pm0.10$ & $3.70\pm0.10$ \\ 
Water content [ppm]  & $36\pm1$  &  $9\pm2$ \\ 
attenuation length at 430 nm [m]  & $7.8\pm0.5$  & $13.5\pm1.0$ \\ 
Light yield [\% BC-505]  &  $48.1\pm0.5$ & $46.6\pm1.0$ \\ 
Potassium [ppb]  & $<2$ & $<2$ \\ 
Refractive index at 405~nm (18$^\circ$) & 1.47 & 1.46 \\ 
Gd-concentration [wt.\%]  & $0.123\pm0.002$ &  -- \\
Hydrogen fraction [wt.\%]  & $13.60\pm0.04$ (calculated) &  $14.6\pm0.2$ (measured) \\ \hline \\
\end{tabular}
\end{center}
\end{table}

\section{Conclusion}
A novel Gd-loaded scintillator was produced on the multi-ton (20~m$^3$) scale, all at the laboratory level. This Gd-loaded scintillator (at the level of 10~m$^3$) constitutes the Target liquid in the Double Chooz reactor neutrino oscillation experiment. The design goals for the relevant properties, particularly the challenging optical and radiopurity specifications, were achieved or exceeded. In parallel to the Target, a second scintillator without Gd-loading was designed and produced. Its properties were tuned and optimized for the needs of the experiment. A summary of the scintillator characteristics is listed in table~\ref{Sumprop}. Modular transportable fluid systems and specialized tanks were designed and used to carry out all the fluid operations. A new ``quadracolumn'' technique, using four active inorganic layers was developed and used for purification of the main fluorescent liquid used in the Target and GC scintillator. The scintillators were filled into the Double Chooz detector and are used for reactor antineutrino detection. Stable and high quality scintillator performance was demonstrated during several months of DC data taking~\cite{DC1}.   

\section*{Acknowledgements}
We would like to thank Dr.~Matthias Laubenstein of the Laboratori Nazionale del Gran Sasso and Dr.~Hardy Simgen from MPIK Heidelberg for low-level activity measurements using \mbox{Ge-spectroscopy}. We also thank M.~Hofmann and Dr.~Honghanh Trinhthi from Technische Universit\"at M\"unchen (TUM) for the NAA measurement contribution and for cross-checking our results on the density and viscosity measurements. We acknowledge helpful discussions with Dr.~M.~Neff and P.~Pfahler of TUM as regards chemical region design of the DC detector. We additionally acknowledge important helpful discussions with a number of companies where extra-ordinary discussions allowed us to understand their plant chemistry and achieve modifications so as to reach the optical and radiopurity levels: Koch Industries, Oklahoma, USA (extensive scientific and technical support regarding PXE chemistry); Dixie Chemicals, Texas, USA (PXE); Sensient Technologies, Germany (\mbox{Gd-BDK}) and Perkin-Elmer, Netherlands (``Neutrino Grade'' PPO). Finally we acknowledge the support of EdF Electricite de France, Chooz Nuclear Power Station.

\end{document}